\documentclass[11pt]{article}
\pdfoutput=1 % force pdflatex on arXiv

\usepackage[utf8]{inputenc}
\usepackage[T1]{fontenc}

\usepackage{arxiv}

\usepackage{amsmath}
\usepackage{amssymb}
\usepackage{graphicx}
\usepackage{booktabs}
\usepackage{listings}
\usepackage{subcaption}
\usepackage{enumitem}
\usepackage{natbib}
\usepackage{tikz}
\usetikzlibrary{arrows.meta,positioning,calc,fit,backgrounds,shapes.geometric,decorations.pathreplacing}

\usepackage{hyperref}
\usepackage{url}
\hypersetup{
  colorlinks=true,
  linkcolor=arxivaccent,
  citecolor=arxivaccent,
  urlcolor=arxivaccent,
  filecolor=arxivaccent,
  breaklinks=true,
}

\definecolor{codebg}{rgb}{0.95,0.95,0.97}
\definecolor{codegreen}{rgb}{0.0,0.5,0.0}
\definecolor{codegray}{rgb}{0.5,0.5,0.5}
\definecolor{codepurple}{rgb}{0.58,0,0.82}
\definecolor{codeblue}{rgb}{0.0,0.0,0.7}
\lstdefinestyle{pythonstyle}{
    backgroundcolor=\color{codebg},
    basicstyle=\ttfamily\small,
    breaklines=true,
    captionpos=b,
    commentstyle=\color{codegreen},
    keywordstyle=\color{codeblue}\bfseries,
    numberstyle=\tiny\color{codegray},
    stringstyle=\color{codepurple},
    language=Python,
    showstringspaces=false,
    numbers=left,
    numbersep=5pt,
    frame=single,
    rulecolor=\color{codegray},
    tabsize=4,
}
\title{Metronome: Bound the Cache, Keep the Beat for Real-Time Interaction Model Serving}

\author{%
  \begin{tabular}{c@{\hspace{4em}}c}
    Jiaying Meng & Bojie Li \\
    Independent Researcher & Pine AI
  \end{tabular}
}

\date{}
\runningtitle{Metronome: Bound the Cache, Keep the Beat for Real-Time Interaction Model Serving}

\begin{document}
\maketitle

\begin{abstract}
Real-time interaction models --- Moshi, MiniCPM-o, Qwen-Omni --- turn serving into a \emph{periodic
real-time task}: on every frame a session ingests streaming audio and must respond by a recurring
wall-clock deadline, while its KV cache grows monotonically and stays pinned for the whole conversation.
This regime hides a dangerous failure mode. On a real full-duplex stack, sustained load does not degrade
serving gracefully: it \textbf{falls off a cliff}, jumping in one step from milliseconds per frame to a
stalled engine when accumulated session state exhausts the KV pool. The collapse is \emph{metastable} ---
identical five-minute runs collapse or survive on run-to-run variance --- and \emph{silent}: latency
and deadline-miss metrics read healthy throughout.

We show one move restores both stability and observability: \textbf{bound each
session's resident state, and latency starts telling the truth.} Metronome's in-engine KV window
eliminates the collapse ($0/20$ vs.\ $14/20$ runs across two batches) and turns per-frame latency into a monotone load
signal, on which an online admission controller discovers the schedulable concurrency; without the
window, the identical controller over-admits into the wall. A first-order model predicts the collapse time within a
few percent on the headline model, and a quality probe validates the bound's design by ablation: the
window alone is quality-free in turn-based decoding, and its few pinned attention-sink tokens are what
keep free-running generation healthy. Everything is measured end-to-end on real audio, across four
interaction models on one GPU.
\end{abstract}

\begin{center}
{\footnotesize Code: \url{https://github.com/19PINE-AI/metronome}}\\
{\footnotesize Website: \url{https://01.me/research/metronome}}\\[0.5em]
\includegraphics[width=0.71\linewidth]{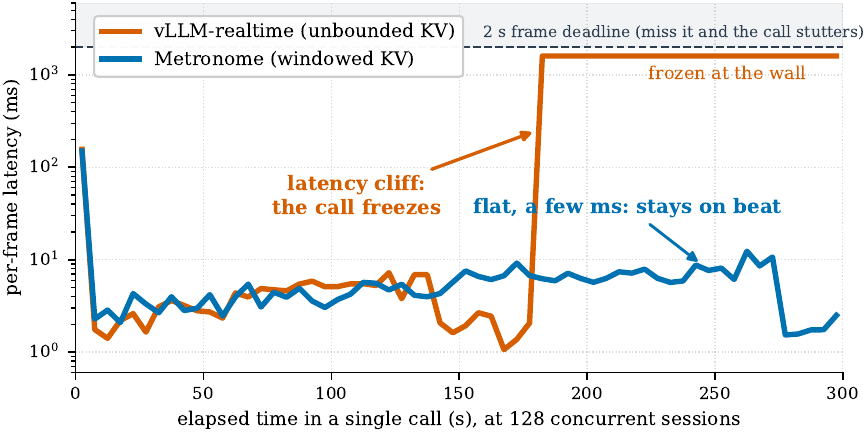}
\vspace{-0.4em}
\captionof{figure}{\textbf{Existing serving falls off a latency cliff; Metronome stays on beat.} One
five-minute call at 128 sessions (Qwen3-Omni-30B): unbounded resident KV jumps in one step from a few ms
to a ${\sim}1.6$\,s ceiling where the stalled engine stops producing tokens; windowed KV stays flat for
the whole call.}
\label{fig:headline}
\end{center}

\section{Introduction}
\label{sec:intro}

A new class of models has quietly created a serving regime of its own. Kyutai Moshi~\citep{moshi},
MiniCPM-o-4.5~\citep{minicpmo}, and the Qwen-Omni family~\citep{qwen2audio,qwen25omni} listen to streaming
audio and respond continuously, with no turn boundary, and a growing line of streaming speech LLMs follows
the same shape~\citep{miniomni,llamaomni}. Frontier labs are converging on it: Thinking Machines Lab's
\emph{interaction models} interleave 200\,ms micro-turns of multimodal input and output on a persistent
GPU sequence~\citep{tmlinteraction}. At the serving layer these models are not chatbots. A chatbot
request is \emph{ephemeral}: a prompt arrives, tokens are generated, the request and its KV cache are
freed. An interaction \emph{session} is a \emph{periodic real-time task}: on every frame --- every 80\,ms
for Moshi, every 1--2\,s for the omni models --- it ingests a new audio chunk as a small prefill and
decodes a short response, against a recurring wall-clock deadline, for a conversation that runs for
minutes (Figure~\ref{fig:workload}).

This regime sits at an unstudied intersection. LLM serving assumes requests are ephemeral: engines
maximize aggregate throughput and reclaim or swap KV between requests~\citep{orca,vllm,clockwork},
parking whatever state must persist out of GPU memory during the idle gaps between turns. Classical
real-time scheduling assumes periodic tasks have \emph{bounded} state~\citep{liulayland}. An
interaction session violates both: its deadline recurs frame after frame, and its per-session KV cache
grows monotonically and stays pinned --- it never leaves on its own. It also has no idle gap in which
to swap that state out or recompute it, the escape hatches turn-based serving relies on
(\S\ref{sec:bg}), so its KV must stay \emph{resident}. The state-of-the-art engines have grown exactly
this mechanism --- vLLM's resumable requests~\citep{vllmrealtime} and SGLang's streaming
sessions~\citep{sglang,tmlinteraction} keep a session's KV live on the GPU across frames ---
but neither addresses the unbounded-state half.

We show that this combination fails in a characteristic and dangerous way. Driving the unmodified
resumable-KV path with real audio through a real full-duplex stack --- WebSocket clients, a Go
gateway, a GPU worker --- we find that sustained load does not degrade interaction serving gracefully. \emph{It falls
off a cliff} (Figure~\ref{fig:headline}): per-frame latency holds at a few milliseconds and then, in a
single step, pins at a ceiling where the stalled engine stops producing tokens. Three properties define
the failure (\S\ref{sec:wall}). It is a \textbf{memory} event: per-frame compute stays cheap to the
last tick, but each session's KV grows every frame until the engine's block pool saturates and the
scheduler stalls. It is \textbf{metastable}: identical five-minute runs collapse or survive on run-to-run
variance in how fast the pool fills. And it is \textbf{silent}: the stalled engine returns empty frames
on time, so latency and deadline-miss metrics read healthy through the collapse --- the call simply goes quiet.

A cliff is not merely a performance bug; it is a \emph{control} bug. Admission, autoscaling, and load
shedding all act on feedback, and a signal that stays flat until the instant of collapse carries no
information to act on. Graceful degradation is therefore not a nicety; it is the \emph{precondition} for
every overload defense one would build. The central finding of this paper is that one move restores both
stability and observability at once: \textbf{bound each session's resident state, and latency starts
telling the truth.} Capping the resident context turns the cliff into a slope: per-frame latency rises
smoothly and monotonically with load, which keeps every session on beat and gives a feedback controller
a signal it can trust.

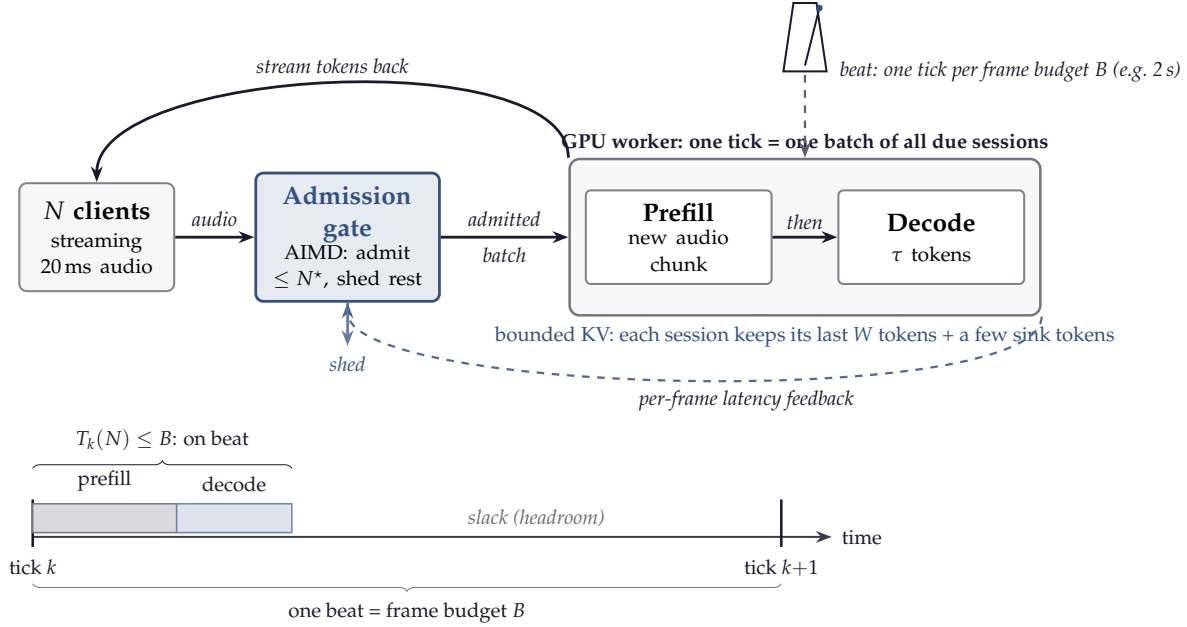
\begin{figure}[t]
\centering
\resizebox{\linewidth}{!}{%
\begin{tikzpicture}[
  font=\small,
  >={Stealth[length=2.2mm]},
  cli/.style={rectangle, rounded corners=3pt, draw=arxivink!55, line width=0.8pt, fill=arxivink!4,
              align=center, inner sep=5pt, text width=18mm, minimum height=15mm},
  gate/.style={rectangle, rounded corners=3pt, draw=arxivaccent, line width=1.1pt, fill=arxivaccent!12,
               align=center, inner sep=5pt, text width=22mm, minimum height=15mm},
  phase/.style={rectangle, rounded corners=2pt, draw=arxivink!55, line width=0.8pt, fill=white,
                align=center, inner sep=4pt, text width=23mm, minimum height=13mm},
  flow/.style={->, line width=1.1pt, draw=arxivink},
  fb/.style={->, line width=0.9pt, draw=arxivaccent!85, dashed},
  lbl/.style={font=\scriptsize\itshape, text=arxivink},
]

% ---- main pipeline: clients -> admission gate -> per-tick batch (prefill -> decode) ----
\node[cli] (cli) {\textbf{$N$ clients}\\[1pt]\scriptsize streaming\\[-2pt]\scriptsize 20\,ms audio};
\node[gate, right=11mm of cli] (gate) {\textbf{\textcolor{arxivaccent}{Admission gate}}\\[-1pt]\scriptsize AIMD: admit\\[-2pt]\scriptsize $\le N^\star$, shed rest};

\node[phase, right=20mm of gate] (pre) {\textbf{Prefill}\\[-1pt]\scriptsize new audio\\[-2pt]\scriptsize chunk};
\node[phase, right=9mm of pre] (dec) {\textbf{Decode}\\[-1pt]\scriptsize $\tau$ tokens};
\draw[flow] (pre) -- node[lbl,above,yshift=-0.3mm]{then} (dec);

% box the two phases into the GPU worker (one tick)
\begin{pgfonlayer}{background}
\node[rectangle, rounded corners=4pt, draw=arxivink!60, line width=1.0pt, fill=arxivink!3,
      fit=(pre)(dec), inner xsep=6pt, inner ysep=12pt] (wk) {};
\end{pgfonlayer}
\node[anchor=south, font=\scriptsize\bfseries, text=arxivink] at (wk.north) {GPU worker: one tick = one batch of all due sessions};
\node[anchor=north, font=\scriptsize, text=arxivaccent] at (wk.south) {bounded KV: each session keeps its last $W$ tokens + a few sink tokens};

% ---- forward arrows ----
\draw[flow] (cli) -- node[lbl,above]{audio} (gate);
\draw[flow] (gate) -- node[lbl,above]{admitted}node[lbl,below]{batch} (wk.west);

% ---- shed (rejected): short arrow straight down from the gate ----
\draw[flow,draw=arxivaccent!75] (gate.south) -- ++(0,-6mm) node[lbl,below,text=arxivaccent]{shed};

% ---- the "beat": a metronome drives one tick per budget ----
\begin{scope}[shift={($(wk.north)+(0,12mm)$)}]
  \draw[arxivink,line width=0.8pt] (-0.30,0)--(0.30,0)--(0.16,0.95)--(-0.16,0.95)--cycle; % body
  \draw[arxivink,line width=0.8pt] (0,0.08)--(0.20,0.88);                                 % pendulum
  \fill[arxivaccent] (0.20,0.88) circle (0.045);                                          % weight
\end{scope}
\node[lbl,anchor=west] at ($(wk.north)+(3.5mm,12mm)$) {beat: one tick per frame budget $B$ (e.g.\ 2\,s)};
\draw[fb,draw=arxivink!70,dashed] ($(wk.north)+(0,11.5mm)$) -- (wk.north);

% ---- outputs back to clients (top arc) ----
\draw[flow] (wk.north west) .. controls ($(wk.north west)+(0,16mm)$) and ($(cli.north)+(0,16mm)$)
   .. node[lbl,above,pos=0.5]{stream tokens back} (cli.north);

% ---- latency feedback to the gate (deep bottom arc, label toward the worker side) ----
\draw[fb] (wk.south east) .. controls ($(wk.south east)+(0,-13mm)$) and ($(gate.south)+(0,-13mm)$)
   .. node[lbl,below,pos=0.45]{per-frame latency feedback} (gate.south);

% ---- anatomy of one tick: the per-frame work must fit inside the beat ----
\begin{scope}[shift={($(cli.south west)+(2mm,-34mm)$)}]
  \def\TW{104mm}\def\PW{20mm}\def\DW{16mm}
  \draw[->,line width=0.8pt,draw=arxivink!85] (0,0) -- (\TW+7mm,0)
      node[anchor=west,font=\scriptsize,text=arxivink]{time};
  \foreach \x/\l in {0/{tick $k$}, \TW/{tick $k{+}1$}}{
    \draw[line width=1.0pt,draw=arxivink] (\x,-1.4mm)--(\x,5.2mm);
    \node[font=\scriptsize,text=arxivink,anchor=north] at (\x,-1.8mm) {\l};}
  \fill[arxivink!16] (0,0.6mm) rectangle (\PW,4.6mm);
  \draw[arxivink!60,line width=0.6pt] (0,0.6mm) rectangle (\PW,4.6mm);
  \fill[arxivaccent!16] (\PW,0.6mm) rectangle (\PW+\DW,4.6mm);
  \draw[arxivaccent!70,line width=0.6pt] (\PW,0.6mm) rectangle (\PW+\DW,4.6mm);
  \node[font=\scriptsize,text=arxivink,anchor=south] at (0.5*\PW,5.0mm) {prefill};
  \node[font=\scriptsize,text=arxivink,anchor=south] at (\PW+0.5*\DW,5.0mm) {decode};
  \node[font=\scriptsize\itshape,text=arxivink!65,anchor=center] at ({0.5*(\PW+\DW+\TW)},2.6mm)
      {slack (headroom)};
  \draw[decorate,decoration={brace,amplitude=3pt},arxivink]
      (0,9.5mm) -- (\PW+\DW,9.5mm)
      node[midway,above=3.5pt,font=\scriptsize,text=arxivink]{$T_k(N)\le B$: on beat};
  \draw[decorate,decoration={brace,mirror,amplitude=3pt},arxivink!60]
      (0,-6.5mm) -- (\TW,-6.5mm)
      node[midway,below=3.5pt,font=\scriptsize,text=arxivink]{one beat = frame budget $B$};
\end{scope}

\end{tikzpicture}}
\caption{\textbf{Metronome keeps the beat.} A metronome \emph{tick} fires once per frame budget $B$. On
each tick the system serves one \emph{batch} of all due sessions --- prefill each session's new audio
chunk, then decode a few tokens (continuous batching) --- and must finish within $B$; the timeline at the
bottom shows one tick's anatomy, with the slack as schedulability headroom. Two interdependent mechanisms
keep the beat steady: (a)~an \emph{admission gate} (AIMD, in the gateway) admits up to the schedulable
concurrency $N^\star$ and sheds the surplus, driven by the per-frame latency it measures; (b)~an
\emph{in-engine windowed KV} (in the worker) retains only each session's last $W$ tokens plus a few
pinned attention-sink tokens, which bounds per-frame work, keeps free-running generation healthy
(\S\ref{sec:quality}), and, by turning the latency cliff into a slope, makes that feedback signal
trustworthy.}
\label{fig:arch}
\end{figure}

\paragraph{Metronome.} We realize this principle as Metronome (Figure~\ref{fig:arch}) and keep the
mechanism deliberately minimal, because the contribution is knowing \emph{what} to bound and \emph{why},
not machinery. An \emph{in-engine windowed KV} activates the engine's dormant sliding-window path so
that each session attends over --- and, crucially, retains --- only its most recent $W$ tokens plus a
few pinned attention-sink tokens that keep a full-attention backbone generating cleanly
(\S\ref{sec:windowed}). On the latency signal this restores, an online
AIMD \emph{admission controller} discovers the schedulable concurrency $N^\star$ and sheds the surplus
cleanly (\S\ref{sec:admission}). The dependency between the two is demonstrated, not asserted: the
identical controller converges on bounded state and over-admits straight into the wall without it
(\S\ref{sec:eval-admission}). A first-order model of pool fill makes the collapse predictable, not
just observable (\S\ref{sec:model}).

% Metronome architecture (intro): the "beat" (one tick per frame budget) drives per-tick BATCH serving.
% Shows (a) admission gate feeding the batch, (b) the prefill->decode phases inside a tick, (c) windowed KV.

\paragraph{Contributions.}
\begin{enumerate}[leftmargin=1.4em,itemsep=2pt]
\item \textbf{Characterization.} Real-time interaction serving is periodic serving with unbounded
per-session state. On a real full-duplex stack, we show that this combination fails by a
\emph{memory-triggered, metastable, silent} latency cliff, and we give a validated first-order model
that predicts when it strikes and explains its apparent randomness
(Sections~\ref{sec:wall},~\ref{sec:model}).
\item \textbf{Principle.} Bounding each session's resident KV removes the cliff \emph{and} restores a
monotone latency signal. One mechanism buys both stability and observability, and the second is what any
overload control requires (\S\ref{sec:metronome}).
\item \textbf{Demonstration.} A minimal in-engine bound --- sliding window plus pinned attention
sinks --- and a deadline-aware admission controller that demonstrably depends on it, measured
end-to-end on real audio across four interaction models, with each half's role in generation quality
validated by ablation (\S\ref{sec:eval}).
\end{enumerate}

\section{Interaction Sessions Are Periodic Real-Time Tasks}
\label{sec:bg}

% The workload contrast (Sec. 2): a chatbot/agent request is turn-based (KV parked out of
% GPU memory between turns -- swapped or recomputed -- with a reactivation toll that grows);
% an interaction session is persistent and periodic (small work every frame, no idle gap,
% resident KV grows monotonically and stays pinned for the whole conversation).
\begin{figure}[t]
\centering
\resizebox{\linewidth}{!}{%
\begin{tikzpicture}[
  font=\small,
  >={Stealth[length=2.0mm]},
  lbl/.style={font=\scriptsize\itshape, text=arxivink},
  rowlbl/.style={font=\small\bfseries, text=arxivink, anchor=east, align=right},
]
\def\TL{118mm}   % timeline length

% =============== row (a): chatbot / agent request (turn-based) ===============
\begin{scope}[shift={(0,0)}]
  \node[rowlbl, text width=24mm] at (-3mm, 4mm) {chatbot /\\[-2pt]agent request};
  \draw[->, line width=0.8pt, draw=arxivink!85] (0,0) -- (\TL+5mm,0)
      node[anchor=west, font=\scriptsize, text=arxivink]{time};
  % turn 1: prefill + decode burst
  \fill[arxivink!16]    (2mm,0.8mm)  rectangle (10mm,6.2mm);
  \draw[arxivink!60, line width=0.6pt]  (2mm,0.8mm)  rectangle (10mm,6.2mm);
  \fill[arxivaccent!16] (10mm,0.8mm) rectangle (26mm,6.2mm);
  \draw[arxivaccent!70, line width=0.6pt] (10mm,0.8mm) rectangle (26mm,6.2mm);
  \node[font=\scriptsize, text=arxivink, anchor=south] at (6mm,6.4mm) {prefill};
  \node[font=\scriptsize, text=arxivink, anchor=south] at (18mm,6.4mm) {decode};
  % idle gap
  \node[lbl, anchor=center, text=arxivink!70] at (44mm,3.4mm) {idle: user thinks, tool runs\ldots};
  % turn 2: bigger re-prefill + decode
  \fill[arxivink!16]    (62mm,0.8mm) rectangle (74mm,6.2mm);
  \draw[arxivink!60, line width=0.6pt]  (62mm,0.8mm) rectangle (74mm,6.2mm);
  \fill[arxivaccent!16] (74mm,0.8mm) rectangle (92mm,6.2mm);
  \draw[arxivaccent!70, line width=0.6pt] (74mm,0.8mm) rectangle (92mm,6.2mm);
  \node[font=\scriptsize, text=arxivink, anchor=south] at (77mm,6.4mm) {turn 2};
  % KV footprint sawtooth: rises in turn 1, parked between turns, restored + grows in turn 2
  \draw[line width=1.2pt, draw=arxivink!70]
      (2mm,-9mm) -- (26mm,-5.4mm) -- (26mm,-9mm) -- (60mm,-9mm) -- (60mm,-6.2mm)
      -- (62mm,-6.2mm) -- (92mm,-4.2mm) -- (92mm,-9mm) -- (\TL,-9mm);
  \node[font=\scriptsize, text=arxivink!70, anchor=east, rotate=90] at (-0.5mm,-7mm) {KV};
  % swap out / swap in arrows to the host band
  \draw[->, line width=0.9pt, draw=arxivink!60] (28mm,-5.6mm) -- (28mm,-12mm);
  \draw[->, line width=0.9pt, draw=arxivink!60] (57mm,-12mm) -- (57mm,-6.6mm);
  \node[font=\scriptsize, text=arxivink!85, anchor=east] at (27mm,-11mm) {swap out};
  \node[font=\scriptsize, text=arxivink!85, anchor=west] at (58mm,-11mm) {swap in};
  % host memory band
  \draw[line width=0.6pt, draw=arxivink!35, dashed] (2mm,-13mm) -- (\TL,-13mm);
  \node[lbl, anchor=west, text=arxivink!70] at (2mm,-15.8mm)
      {host DRAM across PCIe (or drop and recompute) --- movement hidden inside the idle gaps};
  % reactivation toll annotation
  \node[lbl, anchor=west, text=arxivink] at (94mm,-4.8mm)
      {context accumulates $\rightarrow$};
  \node[lbl, anchor=west, text=arxivink] at (94mm,-7.2mm)
      {each reactivation costs more};
\end{scope}

% =============== row (b): interaction session (persistent, periodic) ===============
\begin{scope}[shift={(0,-32mm)}]
  \node[rowlbl, text width=24mm] at (-3mm, 4mm) {interaction\\[-2pt]session};
  \draw[->, line width=0.8pt, draw=arxivink!85] (0,0) -- (\TL+5mm,0)
      node[anchor=west, font=\scriptsize, text=arxivink]{time};
  % periodic frames: small prefill+decode every period B, for the whole conversation
  \foreach \i in {0,...,7}{
    \pgfmathsetmacro{\x}{2 + \i*14.5}
    \fill[arxivink!16]    (\x mm,0.8mm) rectangle (\x mm + 3.2mm,6.2mm);
    \draw[arxivink!60, line width=0.5pt]  (\x mm,0.8mm) rectangle (\x mm + 3.2mm,6.2mm);
    \fill[arxivaccent!16] (\x mm + 3.2mm,0.8mm) rectangle (\x mm + 5.6mm,6.2mm);
    \draw[arxivaccent!70, line width=0.5pt] (\x mm + 3.2mm,0.8mm) rectangle (\x mm + 5.6mm,6.2mm);
    % recurring deadline tick
    \draw[line width=0.7pt, draw=arxivaccent] (\x mm + 12.5mm,-1.2mm) -- (\x mm + 12.5mm,7.0mm);
  }
  \node[font=\scriptsize, text=arxivink, anchor=west] at (33mm,9.7mm)
      {audio chunk in $\rightarrow$ tokens out, every frame};
  \draw[decorate, decoration={brace, amplitude=2.5pt}, arxivaccent]
      (14.5mm,8.6mm) -- (29mm,8.6mm)
      node[midway, above=2.5pt, font=\scriptsize, text=arxivaccent]{recurring deadline $B$};
  \node[lbl, anchor=west, text=arxivink!75] at (\TL-13mm,10.5mm) {\ldots for minutes};
  % KV footprint staircase: grows monotonically, never freed
  \draw[line width=1.2pt, draw=arxivaccent!85]
      (2mm,-6.5mm) \foreach \i in {0,...,7}{ -- ++(11.0mm,0) -- ++(0,0.3mm) -- ++(3.5mm,0.3mm)};
  \node[lbl, anchor=west, text=arxivaccent] at (40mm,-8.4mm)
      {resident KV grows \emph{every} frame and stays pinned --- it never leaves on its own};
  \node[font=\scriptsize, text=arxivink!70, anchor=east, rotate=90] at (-0.5mm,-4.5mm) {KV};
\end{scope}
\end{tikzpicture}}
\caption{\textbf{The workload engines were not built for.} \textbf{(a)}~A chatbot or agent request is
\emph{turn-based}: bursts of prefill and decode separated by idle gaps. Its context accumulates, so
between turns the engine parks the KV out of GPU memory --- swapped to host, or dropped and
recomputed --- and pays a \emph{reactivation toll} that grows with the conversation; the idle gaps hide
the movement, and a slow reactivation delays only one reply. \textbf{(b)}~An interaction session is
\emph{persistent and periodic}: a small prefill and a short decode against a recurring wall-clock
deadline $B$ on every frame, with no idle gap, for the whole conversation --- its KV must stay resident
(\S\ref{sec:bg}), and resident it grows monotonically, pinned for the session's life. Periodic
deadlines plus unbounded per-session state is the combination this paper studies.}
\label{fig:workload}
\end{figure}
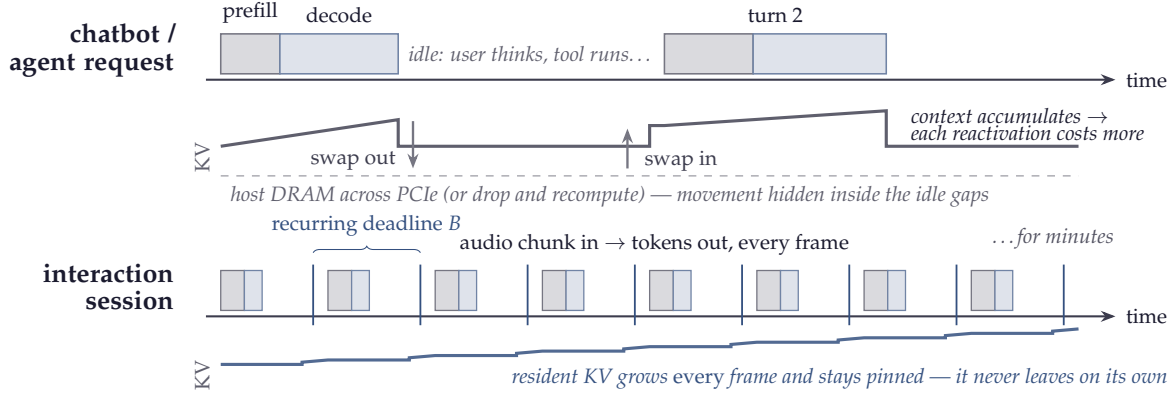

\paragraph{The task model.} A session presents a new audio chunk once per frame; we take the frame
period equal to the frame budget $B$. At frame $k$ the engine must encode and prefill the new chunk, then
decode up to $\tau$ output tokens, attending over the session's accumulated context. With $N$ concurrent
sessions the engine does this for all due sessions each tick under continuous batching~\citep{orca}. The
session is schedulable iff the per-frame wall time satisfies $T_k(N)\le B$ for every $k$ --- a
recurring-deadline condition in the classical real-time sense~\citep{liulayland}. Two properties separate
this from chatbot serving (Figure~\ref{fig:workload}). The deadline \emph{recurs}, so tail behavior
compounds over thousands of frames rather than being amortized away. And the per-session KV cache is
\emph{pinned}: it cannot be reclaimed mid-session the way throughput-oriented swapping assumes, and left
to itself it grows with every frame for the life of the conversation.

\paragraph{Recompute, swap, or stay resident.} Session state that must persist has three places to
live, each taxing a different resource. It can be \emph{recomputed} --- freed, then re-encoded from
recent context on reactivation, echoing sliding-window and streaming
attention~\citep{longformer,mistral,streamingllm} --- a compute toll that grows with the context. It
can be \emph{swapped} to host memory and moved back over the interconnect when next served, as
throughput-oriented engines do between requests~\citep{vllm,flexgen} --- a bandwidth toll that likewise
grows. Or it can stay \emph{resident} in GPU memory: no reactivation toll, but the memory is held.
Turn-based serving can afford the first two because a chatbot's idle gaps --- user think time, tool
calls --- hide the movement. An interaction session has no lull: \emph{every} session is due on
\emph{every} frame, so a recompute or swap toll recurs against the frame budget and grows with session
age --- at our operating points, swapping each due session's state out and back every frame would move
tens of gigabytes per second within minutes. For a periodic real-time session, residency is the only
budget-compatible choice.

\paragraph{What vLLM and SGLang provide, and what they do not.} The state-of-the-art interactive
engines provide exactly this residency. vLLM's resumable-request API~\citep{vllmrealtime} makes a
session one long-lived request: append each frame's chunk, reuse prior KV, no per-frame resubmission
and no movement. SGLang's streaming sessions do the same --- each chunk is appended into a persistent
GPU sequence --- a feature Thinking Machines Lab built and upstreamed to serve its interaction
models~\citep{sglang,tmlinteraction}. This is the correct serving primitive for a session ---
vLLM-realtime is our baseline throughout --- but it \emph{bounds} nothing: the resident context grows
toward \texttt{max\_model\_len} for as long as the session lives, and neither engine offers a notion of
per-frame schedulability or admission. Unbounded state taxes whichever resource it is parked in ---
FLOPs, PCIe, or HBM; residency selects HBM, where the tax accrues silently until the block pool is gone
(\S\ref{sec:wall}). Metronome adds the missing half: bound the resident state.

\section{Anatomy of the Collapse}
\label{sec:wall}

We first dissect how unbounded resident-KV serving fails; everything else in the paper builds on this
diagnosis. Unless noted, the workload is Qwen3-Omni-30B-A3B (FP8)~\citep{qwen3omni} on one Blackwell GPU
behind the real client/gateway/worker stack, with a 2\,s frame budget and $N$ distinct, phase-staggered
real-audio streams (full setup in Section~\ref{sec:method}).

\begin{figure}[t]
\centering
\includegraphics[width=\linewidth]{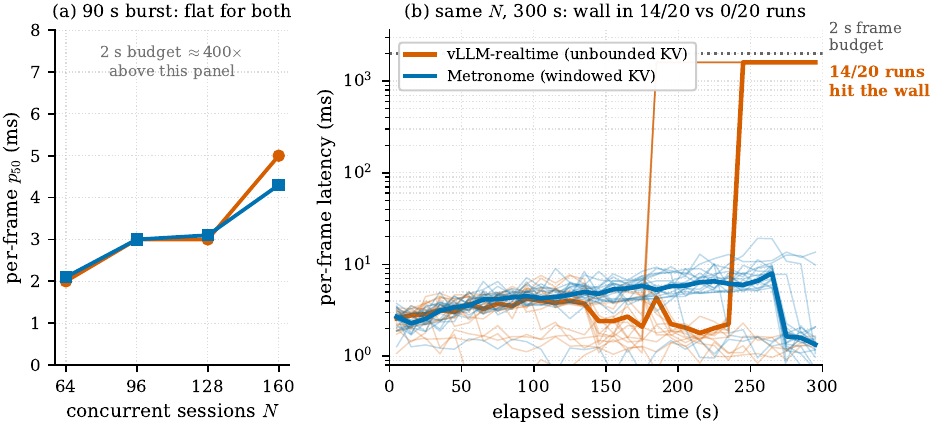}
\caption{\textbf{Bursts are fine; minutes are a gamble.} \textbf{(a)}~Fresh single-$N$ runs over 90\,s:
per-frame latency is flat for \emph{both} policies to $N{=}160$, hundreds of times under the 2\,s
budget --- there is no short-burst capacity problem. \textbf{(b)}~The \emph{same} concurrencies over
300\,s (log scale, 10\,s bucket medians; both twenty-run batches pooled --- ten fresh runs per policy
at each $N{\in}\{96,128\}$; thin lines individual runs, bold the median). With unbounded resident KV,
the block pool saturates, the scheduler stalls, and latency jumps in one step to the ${\sim}1.6$\,s
wall in \textbf{14/20} runs (4/10 on the fixed-order day, 10/10 in the seeded randomized-order
replication); the others end before their pool fills. Windowed KV stays flat in every run
(\textbf{0/20}). The wall is reached by \emph{duration}, not concurrency --- a 90\,s test cannot see
it. Per-run detail in Appendix~\ref{app:results}.}
\label{fig:cliff}
\end{figure}

\paragraph{The failure is a memory cliff, not a compute drift.} Over a 90\,s burst the unbounded baseline
is flawless: latency stays flat far past a hundred concurrent sessions, because every session's resident
context is still small (Figure~\ref{fig:cliff}a). Run the \emph{same} load for five minutes and the shape
changes entirely (Figure~\ref{fig:cliff}b): latency holds at a few milliseconds and then jumps
\emph{discontinuously} to a wall. That binary shape is the tell that the
trigger is not attention compute but memory.

\begin{figure}[t]
\centering
\includegraphics[width=0.92\linewidth]{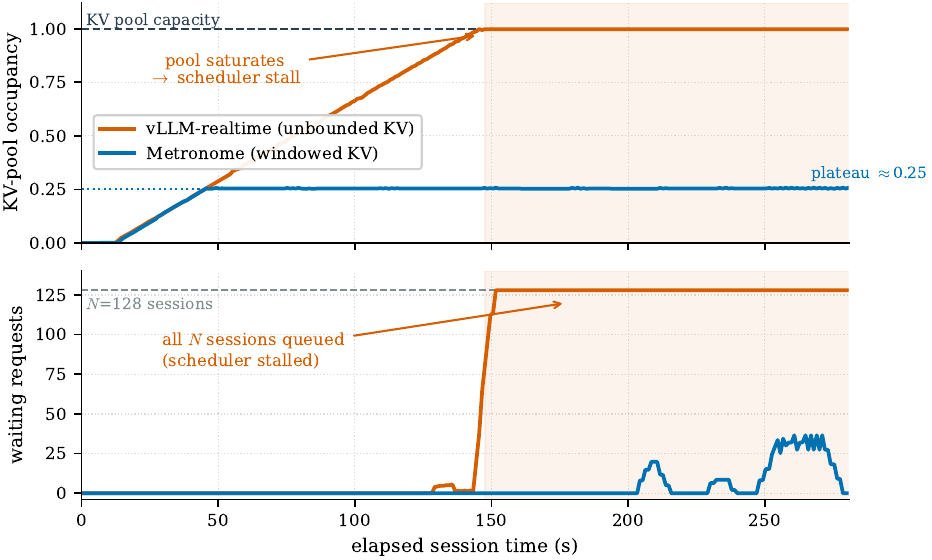}
\caption{\textbf{The trigger, caught in the act} ($N{=}128$, 300\,s, in-engine stat logger).
\textbf{Top:} KV-pool occupancy. Unbounded resident KV climbs monotonically to pool capacity; in-engine
windowed KV frees blocks behind the window and \emph{plateaus} at roughly a quarter of the pool.
\textbf{Bottom:} at the instant of saturation the scheduler can no longer allocate blocks and moves all
$N$ sessions to the waiting queue (shaded), a hard stall that never recovers under open-loop audio. The
windowed policy sees only transient, self-healing preemption late in the run and never approaches
saturation.}
\label{fig:kvpool}
\end{figure}

The mechanism is simple. Each frame appends to every session's resident KV, so $N$ sessions steadily
consume the engine's fixed block pool; when the pool saturates, the scheduler can no longer allocate
blocks and moves every running session to the waiting queue. An in-engine stat logger catches this in the
act (Figure~\ref{fig:kvpool}): pool occupancy climbs monotonically to capacity, at which instant the
running count drops to zero and all $N$ sessions queue. The stall never recovers, because audio keeps
arriving open-loop and the stalled sessions can never catch up.

\paragraph{The collapse is metastable.} Whether a given run collapses is a race between two clocks: the
time the pool takes to fill and the length of the session. At the concurrencies of
Figure~\ref{fig:cliff}b the two are comparable, so run-to-run variance in fill rate --- output length,
prefix-cache sharing across streams --- decides the winner: identical five-minute runs collapse in 4/10
cases, while windowed KV never tips. A replication of the same twenty-run matrix in seeded random order
on a different day walled in \emph{every} vanilla run ($10/10$, against $0/10$ windowed): the tip rate
itself moves with the day's fill rate, exactly as a race should, while the asymmetry is the invariant.
The randomness is not inherent: a model that fills its pool much
faster than the session ends collapses \emph{deterministically} --- MiniCPM-o-4.5 reproduces the wall
this way (\S\ref{sec:generality}) --- and Section~\ref{sec:model} makes the race quantitative.

\paragraph{The collapse is silent.} The stalled engine does not miss its deadlines --- it returns
\emph{empty} frames on time. The worker caps how long a tick waits for a session's tokens before
returning (at $0.8B$, which is why the wall in Figure~\ref{fig:cliff}b sits at ${\sim}1.6$\,s rather than
at the budget), so stalled ticks still return under budget and the deadline-miss counter reads zero
throughout the collapse. Per-frame latency, the metric an operator would alarm on, is green until the
step. What the user experiences is not a late frame but a conversation that goes quiet mid-call: a
monitoring stack watching latency and deadline misses --- the natural dashboards for this workload ---
sees a healthy system stop talking.

\subsection{The cliff is predictable}
\label{sec:model}

\begin{figure}[t]
\centering
\includegraphics[width=\linewidth]{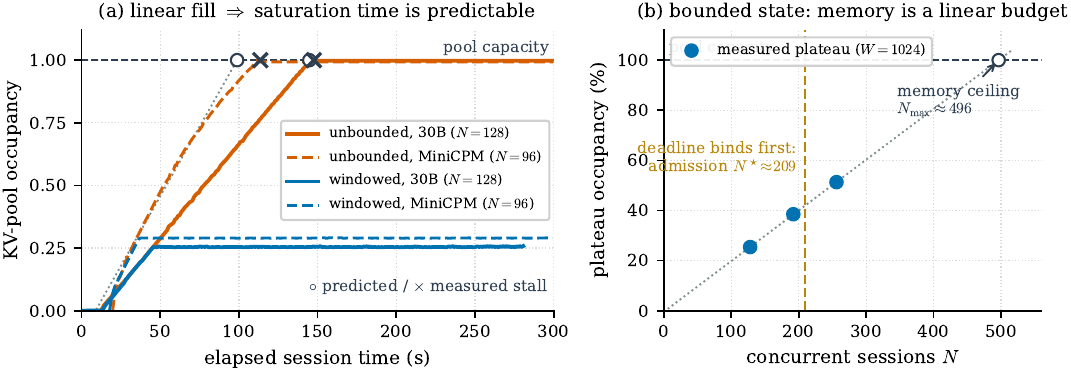}
\caption{\textbf{The collapse obeys a first-order model.} \textbf{(a)}~Unbounded pool occupancy rises
linearly (Eq.~\ref{eq:tsat}); a straight-line fit on the early trace (dotted) predicts the measured
stall ($\times$) within a few percent on Qwen3-Omni-30B ($145$\,s predicted vs.\ $148$\,s) and within
${\sim}13\%$ on MiniCPM-o ($99$\,s vs.\ $114$\,s). Windowed occupancy plateaus far below capacity.
\textbf{(b)}~The windowed plateau is linear in $N$ (${\sim}0.2\%$ of pool per session, $W{=}1024$),
extrapolating to a memory ceiling of ${\approx}500$ sessions --- well above the deadline-schedulable
$N^\star{\approx}209$ that admission discovers, so with bounded state the deadline binds before memory.}
\label{fig:predict}
\end{figure}

The stat-logger traces reveal more than the trigger: they show the collapse obeys a first-order model.
Pool occupancy under unbounded KV rises \emph{linearly} --- each of $N$ sessions appends KV at a steady
per-session rate $r$ (fraction of pool per second), so
\begin{equation}
\rho(t) \;=\; \rho_0 + N r\, t,
\qquad
t_{\mathrm{sat}} \;=\; \frac{1-\rho_0}{N r},
\label{eq:tsat}
\end{equation}
and a run collapses iff $t_{\mathrm{sat}}$ is shorter than the session. Fitting $r$ on the early trace
predicts the measured saturation instant within a few percent on the 30B and within ${\sim}13\%$ on
MiniCPM-o (Figure~\ref{fig:predict}a). The metastability of Section~\ref{sec:wall} now has numbers:
where $t_{\mathrm{sat}}$ is comparable to the session length, ordinary variance in $r$ moves runs across
the boundary, while a faster-filling configuration sits far below it and collapses every time.

Bounding the resident context replaces the linear ramp with a plateau. Each windowed session retains at
most a fixed share of the pool, so occupancy settles at $\rho_\infty = \rho_0 + N s(W)$, linear in $N$
with slope $s(W)$ --- about $0.2\%$ of the pool per session at our operating window
(Figure~\ref{fig:predict}b). Memory turns from a hidden failure clock into a provisionable budget: the
plateau extrapolates to an absolute ceiling of roughly $500$ resident sessions, more than double the
deadline-schedulable concurrency the admission controller discovers
(\S\ref{sec:eval-admission}). Bounded-KV serving is therefore \emph{compute}-limited: the deadline
binds long before memory. This is the exact reverse of the vanilla failure, where memory kills sessions
whose compute the GPU could easily carry.

\section{Metronome: Bound the State}
\label{sec:metronome}

Metronome is one organizing move --- bound each session's resident KV --- and two mechanisms that follow
from it. The \emph{in-engine windowed KV} does the bounding and removes the cliff
(\S\ref{sec:windowed}); the \emph{deadline-aware admission controller} acts on the latency signal the
window makes trustworthy (\S\ref{sec:admission}). Figure~\ref{fig:arch} shows where each sits. The
end-to-end path is deliberately conventional: WebSocket clients stream 20\,ms audio to a Go gateway,
which once per tick issues a single batched \texttt{Step} over gRPC for all due sessions; a GPU worker
owns the engine, feeds each session's long-lived resumable request, and returns the tokens produced
since the last tick. Identical clients and gateway drive every configuration we evaluate.

\subsection{In-engine windowed KV, anchored by sinks}
\label{sec:windowed}

Metronome bounds each session's resident state to a fixed-shape set: the most recent $W$ tokens plus
the session's first few tokens, kept pinned. The \emph{window} bounds both quantities that grow with
session age --- per-frame attention becomes $O(W)$ and resident KV is capped --- and the pinned
\emph{sink} tokens (one--two KV blocks) anchor generation quality: full-attention backbones
concentrate softmax mass on the earliest positions~\citep{streamingllm}, so a bound that evicted them
would corrupt free-running decode. Section~\ref{sec:quality} validates each half by ablation. Per-token
KV-reduction techniques~\citep{mqa,gqa,h2o,scissorhands,snapkv,kivi} shrink what each token stores; we
bound \emph{which} tokens stay resident, and the two compose.

The realization is minimal. vLLM already implements sliding-window attention, including freeing KV
blocks behind the window, but only for models that \emph{declare} a window --- and interaction models do
not, so the path lies dormant. Metronome activates it and completes it: $W$ is installed on the
decoder-attention layers of the omni text backbone at construction time, the KV manager pins each
request's first blocks instead of freeing them, and the attention kernel's windowed mask admits
$[0, S) \cup [t{-}W, t]$ (implementation detail in Appendix~\ref{app:window}). No request
recycling, no re-encode, no application-level machinery: the resident request keeps growing logically
while the engine attends and retains only within the bound. The obvious alternative --- recycling the
resumable request at window boundaries in the application --- achieves the same memory horizon at a
re-encode cost the in-engine bound never pays (\S\ref{sec:eval-cliff}).

\subsection{Deadline-aware admission on a faithful signal}
\label{sec:admission}

Beyond the schedulable concurrency $N^\star$, admitting one more session degrades \emph{every} session,
because continuous batching shares the GPU across all of them; the right behavior is to admit up to
$N^\star$ and shed the rest. But $N^\star$ depends on model, window, hardware, and arrival
pattern, so a hand-set cap is brittle --- it must be \emph{discovered} from feedback. Metronome's
controller is online and model-free: an AIMD loop~\citep{aimd,jacobson} on the measured per-frame
latency, which lowers the admission cap multiplicatively when latency nears a target fraction of the
budget and raises it additively to probe when there is headroom. Arrivals beyond the cap receive a clean
overload rejection rather than degraded service.

This design makes a falsifiable claim: the loop works \emph{only if} latency is a monotone signal of
load. On bounded KV it is, and the controller should converge on $N^\star$. On unbounded KV the signal is
flat until the pool exhausts (\S\ref{sec:wall}), so the same controller should see nothing but headroom
and over-admit into the wall. Section~\ref{sec:eval-admission} runs both arms.

\section{Evaluation}
\label{sec:eval}

We ask four questions, one per subsection: does bounding the state remove the cliff
(\S\ref{sec:eval-cliff})? Does admission converge only on the bounded signal
(\S\ref{sec:eval-admission})? Does the bound cost answer quality (\S\ref{sec:quality})? And does the
failure generalize across models (\S\ref{sec:generality})?

\subsection{Setup}
\label{sec:method}

All experiments run end-to-end --- WebSocket clients, Go gateway, GPU worker --- on one NVIDIA RTX PRO
6000 Blackwell, with real audio (LibriSpeech and spoken-question clips) streamed continuously in 20\,ms
chunks; each session is a distinct, phase-staggered stream, so prefix-cache deduplication cannot
inflate capacity. We report per-frame latency percentiles bucketed by elapsed time, frame-delivery
cadence, and answer correctness under load. Two methodology notes: every data point runs on a freshly
started worker, because a long-lived worker silently inflates apparent degradation
(Appendix~\ref{app:method}); and running these models on Blackwell required four engine bug-fixes,
which ship with our artifact (Appendix~\ref{app:fixes}). The baseline everywhere is unmodified
vLLM-realtime resumable-KV serving --- the strongest available starting point, and the same engine,
stack, and load as Metronome in every comparison.

\subsection{Bounding the state removes the cliff}
\label{sec:eval-cliff}

\paragraph{Headline: $14/20$ walls become $0/20$.} With identical stack, model, and load, differing
only in Metronome's bounded KV, unbounded serving walls in $14/20$ runs pooled across the two
twenty-run batches; windowed serving in $0/20$ (Figures~\ref{fig:cliff}b and~\ref{fig:kvpool};
per-run detail in Appendix~\ref{app:results}). The insurance is free: in runs where the baseline
happens not to collapse, the two policies sit within a few milliseconds of each other.

\paragraph{In-engine beats application-level.} Recycling the resumable request at window boundaries
--- the same memory horizon, implemented above the engine --- also avoids the wall, but its periodic
re-encode grows over the call, while the in-engine window holds flat (Appendix~\ref{app:results}).

\paragraph{The window needs no tuning.} Latency is unchanged across window sizes up to ${\sim}80$\,s
of context (Appendix~\ref{app:results}), and the memory ceiling of \S\ref{sec:model} sits more than
double the schedulable concurrency. With wide margins on both sides, pick a window for quality and
let admission handle the deadline.

\subsection{Admission converges only with bounded state}
\label{sec:eval-admission}

\begin{figure}[t]
\centering
\includegraphics[width=\linewidth]{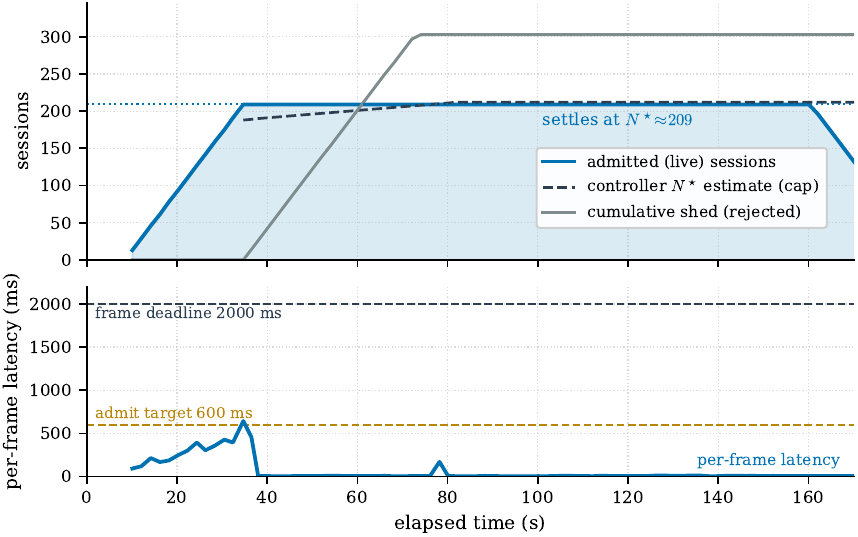}
\caption{\textbf{Online admission discovers the schedulable concurrency --- on the bounded-KV signal}
(open-system ramp: $512$ sessions offered at $8$/s, $600$\,ms latency target, windowed worker).
\textbf{Top:} admitted concurrency and the controller's cap rise with arrivals and settle at
$N^\star{\approx}209$; the surplus is shed cleanly (cumulative count, same axis). \textbf{Bottom:}
per-frame latency --- the only signal the controller uses --- probes the target once during ramp-up,
then holds far under the $2$\,s deadline (steady $p_{99}$ of $12$\,ms). Against unbounded KV the
identical controller admits past this point on a flat signal and ends pinned at the
${\sim}1.6$\,s wall (\S\ref{sec:eval-admission}).}
\label{fig:admission}
\end{figure}

\paragraph{With the bound, the controller converges.} In an open-system overload against the windowed
worker, the cap and the admitted count climb together as sessions arrive; when per-frame latency
first touches the target, the controller caps, and the system settles at $N^\star{\approx}209$ ---
discovered online, with no hand-set capacity number anywhere in the configuration
(Figure~\ref{fig:admission}). Latency holds far under the deadline for the rest of the run; the
surplus is shed cleanly.

\paragraph{Without it, the identical controller over-admits into the wall.} Against unbounded KV the
latency signal stays flat as sessions pour in, so the controller reads pure headroom and admits far
past the limit the windowed system identified, while the admitted sessions' resident KV silently
fills the pool. The run ends pinned at the wall: shedding late cannot rescue sessions that are
already resident. The dependency runs in both directions --- bounded state makes the latency signal
faithful, and a faithful signal makes admission converge.

\subsection{Quality: both halves of the bound, validated by ablation}
\label{sec:quality}

Bounding a full-attention model's context could harm answers, so we ablate each half of the bound in
the two decoding regimes an interaction model runs.

\paragraph{Turn-based decoding: the window alone is quality-free.} In the regime a spoken-QA
deployment runs --- each response an EOS-terminated turn over recent context --- windowed and
unbounded serving are indistinguishable: under load, every session states the correct answer to its
spoken question, with per-frame correctness statistically identical. Each turn's context fits inside
the window; the sinks are not even exercised.

\begin{figure}[t]
\centering
\includegraphics[width=0.9\linewidth]{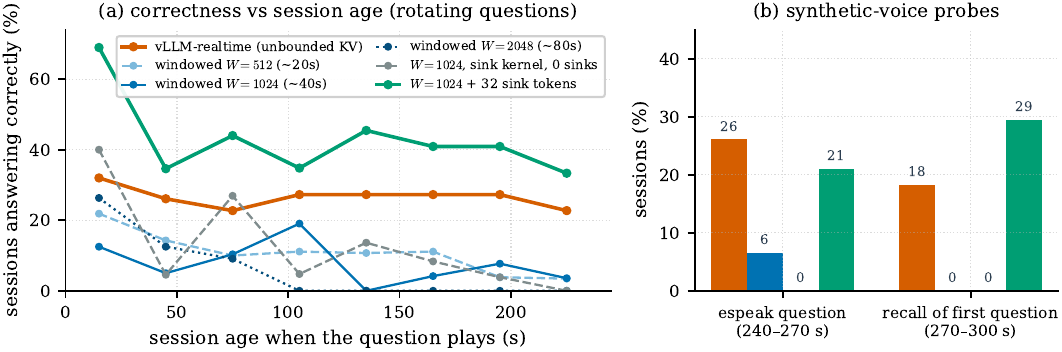}
\caption{\textbf{The bound needs both halves: ablating the sinks fails free-running generation}
(continuous forced decoding, $N{=}32$, 300\,s, Qwen3-Omni-30B; rotating spoken questions, one per 30\,s
segment). \textbf{(a)}~Sessions answering the currently playing question, by session age: unbounded KV
holds a steady low plateau (continuous greedy decoding is degenerate for all policies); every
sink-ablated window declines toward zero as it slides past the session start --- including the
sink-capable kernel run with the sinks off (grey control) --- while the full bound, $W{=}1024$ plus 32
pinned sink tokens (green), holds an age-independent profile at or above the baseline. \textbf{(b)}~A synthesized-voice
comprehension check late in the session and a recall probe about its start: the full bound answers the
fresh question; its nonzero recall \emph{score} is the lenient scorer crediting coherent in-window
answers, not beyond-horizon memory (\S\ref{sec:quality}). Turn-based decoding shows none of this.}
\label{fig:longhz}
\end{figure}

\paragraph{Free-running decode: the sinks are load-bearing.} The harder regime decodes continuously
on one resident request. We play each session a sequence of distinct spoken questions for five
minutes and track whether it answers the one currently playing (Figure~\ref{fig:longhz}). Ablating
the sinks makes every windowed variant decay toward zero once the window slides past the session
start, while the unbounded baseline holds steady; engine metrics stay clean throughout, so the decay
is the model-side attention-sink effect~\citep{streamingllm}, not a serving artifact. Restoring the
sinks removes it: the full bound holds an age-independent profile at or above the baseline and
answers a fresh synthesized-voice question played late in the call. A zero-sink control on the
identical kernel reproduces the decay, pinning the recovery on the sinks; their cost is two KV blocks
per session and unchanged latency.

\paragraph{What sinks do not buy: memory.} Recall beyond the horizon is impossible under any fixed
bound. Asked late for the session's \emph{first} question, a session confidently names the most
recent one it can still see --- a coherent wrong answer, which is what the lenient keyword scorer
credits as the full bound's nonzero recall score in Figure~\ref{fig:longhz}b. Recall is a task for
retrieval, not residency.

\paragraph{Sizing the bound.} A sweep with boundary controls (Table~\ref{tab:sinksweep}) yields one
rule: \emph{pin structure, not content, and keep the window generous}. $S{=}16$ --- one KV block, the
chat header --- works best, and quality falls as the pin reaches into the first turn's content:
pinned content stays semantically live (sessions keep answering the pinned first question minutes
later), and pinning the model's own first output tokens collapses generation into template echo. The
window side is forgiving --- $W{=}2048$ matches $W{=}1024$ --- but sinks cannot rescue a window too
small for the task ($W{=}512$).

\subsection{Generality across four models}
\label{sec:generality}

\begin{figure}[t]
\centering
\includegraphics[width=0.85\linewidth]{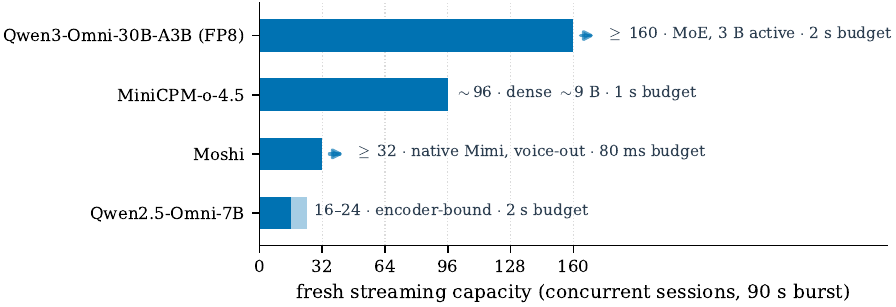}
\caption{\textbf{Fresh single-$N$ streaming capacity across four interaction models} (90\,s burst, one
freshly started worker per point). A $\ge$ arrow means the run was still flat at the largest $N$ tested,
so the value is a lower bound. Short-burst capacity is set by architecture (audio-encoder weight, MoE
sparsity, frame budget); the long-duration cliff of \S\ref{sec:wall} is a property of the serving path
and reproduces across models.}
\label{fig:models}
\end{figure}

\paragraph{Capacity is architectural; the cliff is not.} Fresh single-$N$ capacity spans an order of
magnitude across the four models (Figure~\ref{fig:models}): Qwen2.5-Omni-7B is bound by its
heavyweight audio encoder, while MiniCPM-o-4.5 and the MoE 30B sustain far more sessions on the same
GPU. The cliff, by contrast, travels with the serving path.

\paragraph{On MiniCPM-o, the coin flip becomes certainty.} A dense backbone with 1\,s frames fills its
pool in under two minutes --- a fifth of the ten-minute session, far below the metastable boundary the
30B sat near --- so, exactly as the race of \S\ref{sec:model} predicts, the run hard-stalls on
schedule, deadline-miss counter at zero throughout. Windowed KV under the same load holds
single-digit-millisecond medians for the full ten minutes. Every model on this path has the same KV
growth; the two we drove to the wall differ only in how fast they reach it.

\section{Discussion: Beyond Voice}
\label{sec:discussion}

Nothing in the failure mechanism is specific to audio. Any serving loop that holds \emph{unbounded
per-session state} against a \emph{recurring deadline} manufactures the same cliff: an agent whose
scratchpad and tool-call history grow with every step of a long-running task, a streaming-video
assistant accumulating frame context, a stateful RAG cache that grows per interaction. In each case the
state grows monotonically, is pinned while the session lives, and exhausts a fixed pool on a clock that
per-request latency does not see. The leading indicator is state occupancy, and the principled fix is
the same: bound the resident state per session, and size the bound to the task.

The engine-design implication is that a per-session state bound should be a first-class serving
parameter, not an emergent property of \texttt{max\_model\_len}. Today the model-length cap is the only
backstop on a resident session, and it is a crash boundary rather than a policy: a streaming request that
reaches it does not degrade or stop cleanly --- in our engine it kills every co-resident session at once
(Appendix~\ref{app:fixes}). An engine that exposes ``retain each session's first $S$ and last $W$
tokens, freeing everything between'' as an API --- exactly the sink-anchored bound validated in
\S\ref{sec:quality} --- would give every real-time deployment the slope that control depends on,
without patching model definitions as we do.

\section{Limitations and Future Work}
\label{sec:limits}

\textbf{Scope of hardware and engines.} All results are from one Blackwell GPU and one engine
(vLLM-realtime); the wall is demonstrated on two models (Qwen3-Omni-30B, MiniCPM-o-4.5), admission and
quality on the 30B. A second engine with streaming sessions (e.g.\ SGLang, blocked here by a CUDA
toolchain conflict) would strengthen generality.

\textbf{Statistical coverage.} The wall statistics are the two twenty-run batches of
\S\ref{sec:eval-cliff}; because the tip rate is regime-dependent (\S\ref{sec:wall}), we report the
asymmetry rather than a calibrated probability. Admission, per-model capacity, and the MiniCPM-o wall
results are single fresh runs.

\textbf{Bound-implementation scope.} The full sink-anchored mask lives in the engine's Triton
attention kernel (FlashAttention's window primitive cannot express $[0,S) \cup [t{-}W,t]$); its
quality effect is measured on the 30B at $N{=}32$, across the free-running conditions of
Table~\ref{tab:sinksweep}. The serving-side experiments run the window half alone on the
FlashAttention path --- equivalent for everything they measure, since the sinks add one or two
resident blocks per session and leave latency flat. Replicating the quality boundary on a second
backbone remains open.

\textbf{Workload and modality.} We serve continuous full-duplex streams of real audio; richer
conversational dynamics (turn-taking, barge-in) are out of scope by design, and the omni models emit
text from the thinker --- only Moshi produces streamed voice.

\section{Related Work}
\label{sec:related}

\paragraph{LLM serving engines.} Modern engines optimize aggregate throughput for ephemeral requests via
continuous batching~\citep{orca,vllm}, paged KV~\citep{vllm}, high-throughput single-GPU
execution~\citep{sglang,flexgen}, and multiplexing across models~\citep{alpaserve}; a second line
schedules or splits prefill and decode~\citep{sarathi,distserve,splitwise}. All target ephemeral-request
goodput and are deadline-agnostic, and their state-movement machinery --- reclamation, swapping,
offload~\citep{vllm,flexgen} --- presumes idle gaps that a periodic session never has
(\S\ref{sec:bg}). Metronome treats the resident-KV mechanism these engines now
expose~\citep{vllmrealtime} as its substrate and adds the state bound and deadline-aware control they
lack.

\paragraph{Attention, long context, and KV reduction.} IO-aware exact attention
kernels~\citep{transformer,flashattn,flashattn2,flashinfer} make per-frame attention fast but still scale
with context; sparse and streaming attention bound the cost
structurally~\citep{transformerxl,sparsetransformer,bigbird,longformer,mistral,streamingllm}; per-token
KV-cache reduction~\citep{mqa,gqa,h2o,scissorhands,snapkv,kivi} is orthogonal, shrinking each token's
footprint where our window bounds how many tokens stay resident (\S\ref{sec:windowed}). We repurpose
the engine's sliding-window machinery for the \emph{serving} problem --- bounding resident memory and
per-frame compute on a live streaming session --- and quantify it under a frame deadline, including
the sink anchoring that keeps it quality-free.

\paragraph{Real-time scheduling and admission.} Recurring deadlines and admission control are
classical~\citep{liulayland}, AIMD is the canonical feedback controller~\citep{aimd,jacobson}, and
SLO-aware DNN serving enforces latency targets through predictability and adaptive
batching~\citep{clockwork,clipper}. We instantiate deadline-aware admission online, over a measured
per-frame latency signal, and show the signal itself must be earned by bounding state.

\paragraph{Interaction and speech models.} Full-duplex and streaming speech
dialogue~\citep{moshi,minicpmo,miniomni,llamaomni}, omni understanding~\citep{qwen25omni,qwen2audio,salmonn},
frontier interaction models~\citep{tmlinteraction}, and the encoders and codecs beneath
them~\citep{whisper,audiolm,valle,encodec} define the workload. Our contribution is at the serving
layer, complementary to the models.

\section{Conclusion}
\label{sec:conclusion}

Real-time interaction sessions are periodic real-time tasks with unbounded, pinned per-session state,
and that combination fails in a way dashboards cannot see: not by slowing down, but by a memory-triggered,
metastable, silent cliff. One move fixes more than it appears to. \textbf{Bounding each session's
resident state turns the cliff into a slope}: it keeps every session on beat, makes the collapse
predictable instead of a coin flip, and --- because graceful degradation is what makes a metric
trustworthy --- turns per-frame latency into a signal precise enough to admit exactly to the deadline.
Metronome realizes the principle minimally: a sink-anchored in-engine KV window and the admission
controller it enables, measured end-to-end on real audio across four models, with both halves of the
bound validated by ablation rather than assumed. The lesson generalizes past voice: wherever a real-time loop accumulates unbounded
per-session state, a cliff is being manufactured, and bounding that state is how one earns back the slope
that control depends on.

\section*{Acknowledgements}
This paper was produced using Pine Copilot's voice-directed \emph{whisper coding} workflow~\citep{pineai2026whispercoding}, in which the authors specify, discuss, and review the work by voice while a coding agent, Claude Code with Claude Opus 4.8 and Claude Fable 5, carries out the planning, coding, experiments, and paper writing.
The authors thank BSQL Networking for hosting the NVIDIA RTX PRO 6000 GPU.

\bibliographystyle{plainnat}
\bibliography{reference}

\appendix

\section{In-Engine Windowed KV: Implementation}
\label{app:window}

\paragraph{Window half.} vLLM supports sliding-window attention for models that declare it: when a
decoder layer carries a \texttt{sliding\_window} attribute, the engine constructs a
\texttt{SlidingWindowSpec} for that layer's KV cache and the FlashAttention
backend~\citep{flashattn,flashattn2,flashinfer} applies a windowed causal mask. The KV manager then
frees blocks that fall entirely behind the window, which is what bounds resident memory --- the
property the collapse analysis of Section~\ref{sec:wall} turns on. Metronome sets
\texttt{sliding\_window}${=}W$ at model-construction time on the shared decoder-attention class that
the omni text backbones build, so every decoder layer reports a windowed spec; no scheduler,
allocator, or API change is required.

\paragraph{Sink half.} FlashAttention's window primitive cannot express the sink-anchored mask
$[0,S) \cup [t{-}W,t]$, so the sink half routes the decoder layers to vLLM's Triton unified-attention
backend and extends its kernel: the windowed mask gains the OR-term $k < S$, and the tile loop gains
$\lceil S/\mathrm{tile} \rceil$ extra leading iterations remapped onto the sink tiles --- the online
softmax is order-invariant, so accumulating them first is exact, and in segmented-decode mode only
segment~0 takes them, merged exactly once. On the memory side, the sliding-window KV manager pins each
request's first $\lceil S/\mathrm{block} \rceil$ blocks instead of freeing them, and the per-request
admission cap counts the pinned blocks so pool sizing stays exact. The kernel is verified against a
float32 reference of the union mask (mixed prefill/decode, grouped-query attention, both launch modes),
with freed blocks NaN-poisoned to prove nothing outside $\mathrm{sinks} \cup \mathrm{window}$ is ever
read; with $S{=}0$ every touched path is byte-identical to stock. The test ships with the artifact.

\section{Enabling Omni Streaming on Blackwell}
\label{app:fixes}

Running the omni models on vLLM~0.23 (the first release with resumable requests) on an SM120 Blackwell
GPU required four engine bug-fixes, without which the models do not initialize or stream; the combined
patch that ships with the artifact adds a fifth change, the in-engine bounded KV itself --- the sliding
window and its sink retention (Appendix~\ref{app:window}).

\begin{enumerate}[leftmargin=1.4em,itemsep=2pt]
\item \textbf{cu\_seqlens device.} The shared multimodal encoder attention builds \texttt{cu\_seqlens} on
the feature tensor's device (CPU during memory profiling) while q/k/v are on CUDA, crashing the varlen
flash-attention op~\citep{flashattn}. We coerce it to the query device.
\item \textbf{Capability gate.} FlashInfer~\citep{flashinfer} gates SM12.x support on the \emph{toolkit}
CUDA version. We gate on the \emph{runtime} version so Blackwell is recognized when the local
\texttt{nvcc} is older.
\item \textbf{JIT toolchain.} FlashInfer's bundled CCCL headers reject the CUDA-13.2 \texttt{nvcc}; the
documented escape macro (\texttt{CCCL\_DISABLE\_CTK\_COMPATIBILITY\_CHECK}) unblocks the JIT across a
CUDA minor version.
\item \textbf{mRoPE off-by-one.} A Qwen3-Omni position-id routine double-counts the audio start token,
overshooting the sequence length and crashing precisely when an audio block ends a streamed chunk --- the
per-frame append case. A one-line reconcile fixes it and is a no-op for every previously working input.
\end{enumerate}

With these, Qwen3-Omni-30B, Qwen2.5-Omni-7B, and MiniCPM-o-4.5 all load and stream; Moshi runs on its own
native stack. The evaluation also surfaced a sixth hazard: a resident streaming request that reaches
\texttt{max\_model\_len} does not stop cleanly --- the engine core crashes on a shape mismatch, killing
\emph{every} co-resident session at once. Unbounded per-session state is dangerous at more than one
boundary; our long-horizon runs size \texttt{max\_model\_len} above the session's growth to avoid it, and
Section~\ref{sec:discussion} argues the general fix is a first-class per-session state bound.

\section{Measurement Methodology: Fresh Worker Per Point}
\label{app:method}

Capacity curves for this workload must be measured with a freshly started worker per data point.
Measuring a capacity curve as a \emph{sweep} of many $N$ values on a single long-lived worker silently
inflates the apparent degradation at the larger $N$ values, which arrive later in the sweep: residual
engine state accumulates across points, and slowly varying ambient conditions (in our case, intermittent
contention from an unrelated GPU job) fold into whichever points run last. In our data this made the
unbounded baseline appear to collapse at $N{=}128$--$160$ in a 90\,s sweep, when a fresh worker per point
is flat to $N{=}160$ (Figure~\ref{fig:cliff}a), and it manufactured an apparent capacity cliff for Moshi
at $N{=}24$ that likewise vanished under fresh measurement. Every number in this paper is therefore
measured with one fresh worker per point. We recommend this as standard practice for interaction-serving
capacity measurement: the workload's long-lived resident state makes it unusually sensitive to
measurement-harness state.

\section{Additional Results}
\label{app:results}

\paragraph{Per-run wall outcomes.} Table~\ref{tab:wall} details the per-run outcomes behind
Figure~\ref{fig:cliff}b and \S\ref{sec:eval-cliff}: the fixed-order batch and the seeded
randomized-order replication (the shuffle and per-run logs ship with the artifact).

\begin{table}[h]
\centering\small
\caption{The latency cliff, per operating point (5 fresh runs each, 300\,s; bucket $p_{50}$). Unbounded
resident KV is a metastable gamble whose tip rate moves with the day's fill rate (\S\ref{sec:model});
in-engine windowed KV is flat in every run of both batches.}
\label{tab:wall}
\setlength{\tabcolsep}{4pt}
\begin{tabular}{lcc}
\toprule
& vanilla (unbounded KV) & in-engine windowed KV \\
\midrule
batch 1 (fixed order), $N{=}96$  & wall in $\mathbf{2/5}$ runs ($\rightarrow{\sim}1.6$\,s), else ${\sim}2$--$5$\,ms & ${\sim}1$--$2$\,ms, $\mathbf{0/5}$ wall \\
batch 1 (fixed order), $N{=}128$ & wall in $\mathbf{2/5}$ runs ($\rightarrow{\sim}1.6$\,s), else ${\sim}2$--$5$\,ms & ${\sim}2$--$12$\,ms, $\mathbf{0/5}$ wall \\
batch 2 (randomized), $N{=}96$   & wall in $\mathbf{5/5}$ runs ($\rightarrow{\sim}1.6$\,s) & flat few ms, $\mathbf{0/5}$ wall \\
batch 2 (randomized), $N{=}128$  & wall in $\mathbf{5/5}$ runs ($\rightarrow{\sim}1.6$\,s) & flat few ms, $\mathbf{0/5}$ wall \\
\bottomrule
\end{tabular}
\end{table}

\begin{figure}[h]
\centering
\includegraphics[width=0.62\linewidth]{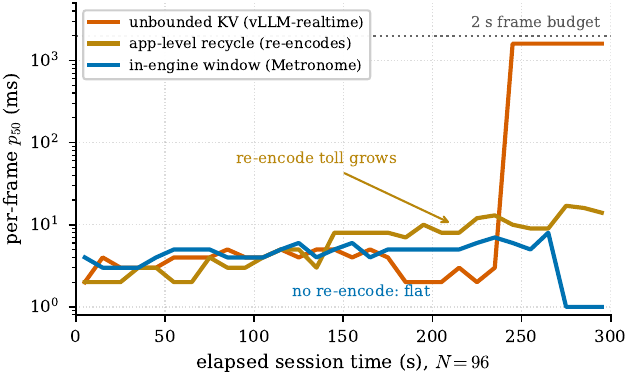}
\caption{\textbf{In-engine windowing beats application-level recycling.} Three policies with the same
memory horizon at $N{=}96$ over 300\,s (per-frame $p_{50}$, 10\,s buckets). Unbounded resident KV hits
the wall at ${\sim}240$\,s. The application-level recycling proxy avoids the wall but its periodic
re-encode grows over the call ($p_{50}$ to ${\sim}14$--$17$\,ms, $p_{90}$ to $36$\,ms). The in-engine
window bounds the horizon without ever re-encoding.}
\label{fig:reencode}
\end{figure}

\paragraph{In-engine windowing vs.\ application-level recycling.} Figure~\ref{fig:reencode} details the
comparison summarized in \S\ref{sec:eval-cliff}: three policies with the same memory horizon.

\begin{figure}[h]
\centering
\includegraphics[width=0.6\linewidth]{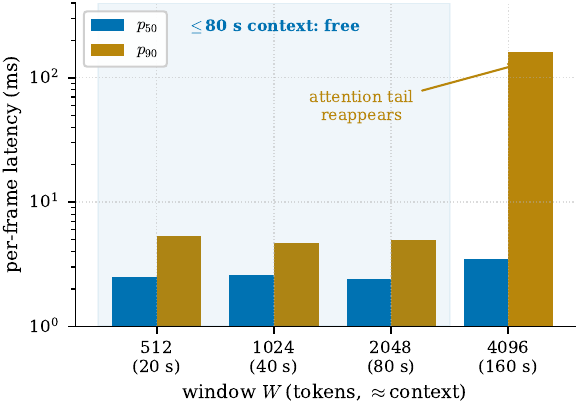}
\caption{\textbf{A generous window is free.} Window-size ablation (in-engine windowed KV, $N{=}128$,
60\,s runs): $p_{50}/p_{90}$ stay ${\sim}5$\,ms up to a 2048-token (${\sim}80$\,s) window.}
\label{fig:wabl}
\end{figure}

\paragraph{Window-size ablation.} Figure~\ref{fig:wabl} sweeps the window size at fixed load. Latency is
flat up to a 2048-token (${\sim}80$\,s) window; the lone $p_{90}$ uptick at 4096 tokens carries a
$p_{99}$ comparable to the smaller windows', so we attribute it to single-run tail noise rather than a
clean attention effect. A memory horizon well beyond the ${\sim}30$\,s most interactions need is free.

\paragraph{Turn-based quality detail.} In the turn-based probe of Section~\ref{sec:quality} ($N{=}96$,
75\,s sessions, EOS-terminated turns), answer-stated counts are $96/96$ sessions for both vanilla and
windowed policies, with per-frame correctness ${\sim}70\%$ vs.\ ${\sim}68\%$ (statistically
indistinguishable). The gap from $100\%$ per-frame is genuine model behavior on looped, sometimes
partial audio and is identical across policies.

\paragraph{Free-running conditions: sweep and boundary controls.} Table~\ref{tab:sinksweep} collects
every free-running condition --- the ablation of Figure~\ref{fig:longhz} plus the sink/window sweep,
with boundary controls at the exact token-layout edges. The controls localize the harm of
over-pinning (\S\ref{sec:quality}): a pin that \emph{truncates} the first audio block is harmless
($S{=}32$), refuting the malformed-block hypothesis, and the failure signatures in the outcome column
track how much first-turn \emph{content} the pin keeps semantically live. Engine-side, the rows
differ only as block arithmetic dictates (pool column: plateau $\propto W$, plus ${\sim}0.1\%$ of the
pool per pinned sink block) and latency is flat in every run, so the quality differences are entirely
model behavior. The unbounded baseline's pool was still climbing when its 300\,s session ended --- at
$N{=}32$ the session outlives the run, the benign side of the two-clock race of \S\ref{sec:model}.

\begin{table}[h]
\centering\small
\caption{All free-running probe conditions (Figure~\ref{fig:longhz} and the sink/window sweep;
$N{=}32$, 300\,s, one fresh worker per row, per-frame $p_{99}{\le}3$\,ms and zero errors throughout).
\emph{Mid-call}: sessions answering the currently playing question over ages 30--240\,s.
\emph{Fresh Q}: the synthesized-voice question at 240--270\,s. \emph{Pool}: KV-pool occupancy
plateau. Token layout: chat header $[0,14)$, first audio $[14,42)$, instruction $[42,53)$,
assistant-open $[53,58)$, generated from $58$.}
\label{tab:sinksweep}
\footnotesize
\setlength{\tabcolsep}{4pt}
\begin{tabular}{llcccl}
\toprule
$(W,\,S)$ & pin covers & mid-call & fresh Q & pool & outcome \\
\midrule
unbounded      & ---                              & $23$--$27\%$ & $26\%$ & $74\%\,\uparrow$ & steady; pool still filling at 300\,s \\
\midrule
$(512,\,0)$    & --- (sinks ablated)              & $\to 3\%$    & $7\%$  & $3.3\%$  & decays as window slides \\
$(1024,\,0)$   & --- (sinks ablated)              & $\to 0$--$8\%$ & $6\%$ & $6.4\%$  & decays \\
$(2048,\,0)$   & --- (sinks ablated)              & $\to 0\%$    & $0\%$  & $12.7\%$ & decays \\
$(1024,\,0)$   & --- (sink kernel, control)       & $\to 0\%$    & $0\%$  & $6.5\%$  & decays: kernel is not the cause \\
\midrule
$(1024,\,16)$  & chat header                      & $38$--$57\%$ & $45\%$ & $6.5\%$  & \textbf{recovers --- best} \\
$(1024,\,32)$  & $+\,{\sim}2/3$ of first audio    & $33$--$45\%$ & $21\%$ & $6.5\%$  & recovers \\
$(1024,\,42)$  & $+$ complete first audio         & $20$--$32\%$ & $21\%$ & $6.6\%$  & degraded: answers the pinned clip \\
$(1024,\,58)$  & $+$ instruction, assistant-open  & $25$--$37\%$ & $18\%$ & $6.8\%$  & degraded: quotes the pinned clip \\
$(1024,\,64)$  & $+$ first 6 \emph{generated} tokens & $\to 7\%$ & $8\%$  & $6.6\%$  & collapses: template echo \\
\midrule
$(512,\,32)$   & as $(1024,32)$                   & $\to 0\%$    & $0\%$  & $3.5\%$  & too-small window; sinks no rescue \\
$(2048,\,32)$  & as $(1024,32)$                   & $40$--$57\%$ & $42\%$ & $12.8\%$ & recovers; matches $W{=}1024$ \\
\bottomrule
\end{tabular}
\end{table}

\end{document}